\documentclass{aastex}

\lefthead{Geisler et al.}
\righthead{\lq Sculptor'-ing the Galaxy?}

\def\feh{\ {[Fe/H]}\ }
\def\met{\ {metallicity}\ }
\def\mets{\ {metallicities}\ }

\def\mp{\ {metal-poor}\ }
\def\mr{\ {metal-rich}\ }

\def\pops{\ {populations}\ }
\def\gtsim{\ {\raise-0.5ex\hbox{$\buildrel>\over\sim$}}\ }
\def\ltsim{\ {\raise-0.5ex\hbox{$\buildrel<\over\sim$}}\ }
\def\alp{\ {$\alpha$}\ }
\def\Glx{\ {Galaxy}\ }
\def\Gal{\ {Galactic}\ }

\def\abu{\ {abundance}\ }
\def\abus{\ {abundances}\ }

\begin{document}

\title{\lq Sculptor'-ing the Galaxy? \\
The Chemical Compositions of Red Giants in the  \\
Sculptor Dwarf Spheroidal Galaxy}

\author{Doug Geisler}
\affil{Departamento de Fisica, Universidad de Concepci\'on, Casilla 160-C,
Concepci\'on, Chile; doug@kukita.cfm.udec.cl}

\author{Verne V. Smith}
\affil{Department of Physics, University of Texas El Paso, El Paso, TX 79968 
USA; verne@barium.physics.utep.edu}

\author{George Wallerstein}
\affil{Astronomy Department, University of Washington, Seattle, WA 98195
USA; wall@orca.astro.washington.edu}

\author{Guillermo Gonzalez}
\affil{Department of Physics \& Astronomy, Iowa State University, Ames, 
IA 50011-3160 USA; gonzog@iastate.edu}

\author{Corinne Charbonnel} 
\affil{Geneva Observatory, CH-1290 Sauverny, Switzerland; Corinne.Charbonnel@obs.unige.ch}
\affil{Laboratoire d'Astrophysique de l'Observatoire Midi-Pyr\'en\'ees, 14 Avenue Edouard Belin, Toulouse, 
F-31400, France}

\begin{abstract}

We  have used high-resolution, high signal-to-noise spectra 
obtained with the VLT and UVES to determine abundances of 17 elements 
in 4 red giants
in the Sculptor dwarf spheroidal galaxy.
Our [Fe/H] values range from --2.10 to --0.97, confirming previous findings of
a large metallicity spread. We have combined our data with 
similar data for five Sculptor giants studied recently
by Shetrone et al. to form one of the
largest samples of high resolution abundances yet obtained for a dwarf
spheroidal galaxy, covering essentially the full known metallicity range in 
this galaxy. These
properties allow us to establish trends of [X/Fe]
with [Fe/H] for many elements, X. The trends are significantly different
from the trends seen in galactic halo and globular cluster stars.
This conclusion is evident for most of the elements from oxygen to manganese.
We compare our Sculptor
sample to their most similar Galactic counterparts
and find substantial differences remain even with these stars.
The many discrepancies in the relationships between [X/Fe] as seen in Sculptor
compared with Galactic field stars indicates that our halo
cannot be made up in bulk
of stars similar to those presently seen in dwarf spheroidal galaxies
like Sculptor, corroborating similar
conclusions       reached by Shetrone et al., Fulbright and Tolstoy et al. 
These results
have serious implications
for the Searle-Zinn and hierarchical galaxy formation scenarios.
We also find that the most metal-rich star in our sample is 
a heavy element-rich star. This star and the [Ba/Eu]
trend we see indicates that AGB stars must have played an important role in 
the evolution of the s-process elements in Sculptor. A very               high
percentage of such heavy element stars are now known in dwarf spheroidals 
compared to the halo, 
further mitigating against the formation of the halo from such objects.

\end{abstract}


\clearpage

\section{Introduction}

The complex relationships among the nearby dwarf spheroidal (dSph) galaxies, the
globular clusters and the general halo of our galaxy are far from clear,
though all belong to Baade's (1944) original Population II.  It is of interest
to discuss the similarities and differences of their stellar populations in
terms of possible scenarios for
the origin of the Galactic halo.  According to Eggen, Lynden-Bell
\& Sandage (1962) the Galactic halo, including the globular clusters,
formed in a single monolithic
event of gravitational collapse lasting about $2\times10^8$
years.  During that time, the vast majority of halo stars and stars in the
globulars formed to provide the present population II.  An alternative
scenario was suggested by Searle \& Zinn (1978) in which the halo was
accumulated by the capture of many small systems such as dSphs
over a timescale at least an order of magnitude longer.
Many current versions of hierarchical galaxy formation theories invoke
similar scenarios (e.g. Klypin et al. 1999, Moore et al. 1999). Strong observational
evidence in favor of at least some contribution of dSph systems
to the halo was first provided by Ibata, Gilmore \& Irwin (1994) with
the discovery that the Galaxy is currently 
capturing the Sagittarius system (Sgr)
with its attendant globular clusters.
More recent evidence (e.g.
Yanny et al. 2003, Martin et al. 2004) indicates that the Galaxy may well be
absorbing or has absorbed additional systems.
In addition, the complicated history of gradual
metal growth in $\omega$ Centauri indicates that it probably orbited our
galaxy for several Gyr before being captured (Hughes \& 
Wallerstein 2000; Hilker \& Richtler 2000).  That the Galaxy, as well as M31
and presumably other similar spirals, are intimately
surrounded by a number of dSphs is
made particularly clear in the graphic
3-dimensional galaxy distribution map of the
Local Group given in Grebel (1999). A full review of current
ideas regarding the formation of the Galaxy can be found in       Freeman
\& Bland-Hawthorn (2002).

One approach that helps to constrain
formation scenarios
is to compare the populations of
the surviving dSph systems with that of the halo. If the halo is indeed made
up in large part by dissolved systems initially
like the dSphs we see today, one would expect
to find many similarities in their stellar \pops. Of various methods of
comparison, three stand out as potentially the most
viable.  The first is a comparison
of the types and period distributions of variable stars (Renzini 1980).
This method is
not currently useable because of selection effects that plague any census 
of variables in the halo, although various surveys such as
the Sloan Digital Sky Survey
should help pin these down.
The second is to compare the CMDs in detail and try to  set limits on the 
percentage of present day
dSph populations that may have contributed to the halo. By
comparing the turnoff colors in these systems, Unavane, Wyse \& Gilmore (1996)
have set an upper limit of $\sim 10\%$ on this contribution, as the 
intermediate-age stars generally found in dSphs are lacking in the halo,
hinting that dSphs may not be the generic galactic building blocks they are
often imagined to be.

The third approach is a direct
comparison of the detailed chemical compositions of 
stars  from the two environments, based on high resolution spectroscopy.
A large sample of Galactic field stars  
with detailed
abundance analyses is now available (e.g. McWilliam 1997; 
Burris et al. 2000; Ryan, Norris \& Beers 1996, Carretta et al. 2002, Nissen \&
Schuster 1997, Fulbright 2002, Johnson 2002, Gratton et al. 2003a, 
Stephens and Boesgaard 2002).
A kinematic analysis of these
and other Galactic field stars shows most belong to the halo below
[Fe/H] = -1, although several thick disk stars have metallicities
that extend below [Fe/H] = -2 (Venn et al. 2004).   Here, we assume
that stars with [Fe/H] below -1 are representative of the Galactic halo.
The complementary studies of stars in dSph systems have only recently begun.
In a pioneering study, Shetrone, Bolte \& Stetson (1998) investigated four stars
in the Draco dSph. They                                        
found a metallicity range from [Fe/H] = $-3.0$ to $-1.5$ with a mean value
of [$\alpha$/Fe] of +0.2 and a spread in [O/Fe] from +0.38 to
$-0.32$.  
Subsequently, Shetrone, C\^ot\'e, \&
Sargent (2001 - hereafter S01) 
published results for an additional two stars in Draco, 
six in Ursa Minor and 5 in Sextans, and Shetrone et al. (2003 - hereafter S03) 
added five
giants each in Carina and Sculptor (hereafter Scl), 
three in Fornax and two in Leo I. 
Combining their studies, they find that at a given \met the dSph giants 
exhibit significantly lower [\alp/Fe] abundance ratios than stars in the
Galactic halo. They conclude that the general \mp Galactic
halo could not be built up 
of stars like those seen in their dSph samples. However, they find that a
small subset of stars, represented by the \mr, high $R_{max}$, high $z_{max}$
halo stars studied by Nissen \& Schuster (1997), do mimic the dSph stars of
similar \met in their detailed chemical composition
and support Nissen \& Schuster's claim that up to 50\% of the \mr
halo could be explained by dSph accretion, although this relied on only two
dSph stars of the appropriate \met.
However, Venn et al. (2004) further analysed the Nissen \& Schuster (1997)
stars and find that the Ni-Na relationship, the basis for the S03
claim about 50\% of the metal-rich halo possibly being stripped dSphs, is
a general nucleosynthetic signature and not relevant to the discussion
of merged galaxies.
Finally, Bonifacio et al. (2000), Bonifacio and Caffau (2003), Bonifacio et al.
(2004) and Smecker-Hane \& McWilliam (2002) have 
studied a large sample of stars in the Sgr dSph. The latter find 
significant
differences between Sgr and Galactic field stars of comparable \met, in particular
with regards to Al, Na and \alp elements, in the same sense as for the samples
of Shetrone and collaborators.
Bonifacio et al. (2004)  suggest that the chemical similarities of dSphs and 
damped Ly\alp systems, particularly in regard to their depressed \alp \abus,
may demonstrate a common evolutionary history and nature.

Gratton et al (2003a) have used kinematics to
divide field subdwarfs and early subgiants into two subpopulations,
one which
they ascribe to a dissipational collapse and one that is likely to represent
accreted stars. In particular, they have compared the ratios of alpha-elements
to iron in the two subpopulations. They find that on average the supposed 
accreted population has lower [\alp/Fe] than their dissipative collapse
counterparts.
If the accreted stars have been
accumulated by the capture of systems like Scl, their compositions should
be similar to those of Scl. 

To expand the data base for dSph galaxies and to further test the hypothesis
that the halo of the Milky Way may have been ``sculpted" from galaxies like Scl,
we have observed 4 red giants in the
Scl galaxy.  Scl was one of the first two dSph companions to the
Milky Way to be discovered (Shapley 1938).  Indeed, as the prototype, dSph
galaxies were originally referred to as ``Scl-like systems" (e.g. Shapley 1943).
Photometry in
Scl (e.g. Da Costa 1984, Schweitzer et al. 1995) 
has shown that the  red
giant branch is broad, indicating a spread in metallicity (as is seen in
all dSphs and in $\omega$ Cen).
In a survey of 37 red giants in Scl, Tolstoy et al. (2001) used the IR 
CaII triplet to derive metallicities\footnote{The estimation of 
metallicities by measurement of the CaII triplet has proven to be very 
efficient and useful.  However, many authors refer to their results as [Fe/H] rather 
than [Ca/H] which was actually measured. The translation from [Ca/H] to \feh
can be problematical (e.g. Cole et al. 2000), especially when [Ca/Fe] may vary
between program objects and abundance calibrators.} ranging from $-0.8$ down to 
$-2.3$ dex with most of the stars distributed between $-1.3$ 
and $-2.0$.  The distribution is asymmetric with a mean of   $-1.5\pm 0.3$ and a
peak near $-1.3$.  Both the mean and spread are very similar to $\omega$ Cen 
(Norris, Freeman \& Mighell 1996; Sunzeff \& Kraft 1996) but the 
asymmetry in $\omega$ Cen is opposite to that of Scl with the peak of the 
distribution on the metal-poor side of the mean.          Apparently the 
two systems had slightly different rates of self-enrichment and star formation.

Scl provides an excellent 
opportunity to compare the compositions of dSph stars
with those in the halo.  At a distance of 87 kpc (Mateo 1998), it is one of the 
nearest dSphs. Its brightest giants, at $V\sim 17.3$, are therefore accessible to
8m-class telescopes equipped with
high resolution spectrographs. Since Scl shows a
spread in metallicity, there is the possibility to compare
the ratios of various species to iron as a function of metallicity with 
the same elemental ratios in the halo field and the
globular clusters. Additionally, detailed abundances are required to disentangle
age from \met effects when attempting to derive the star formation and chemical
enrichment history of Scl, as discussed
by Tolstoy et al. (2003) for their age determinations.
Photometric data alone are not sufficient to break
the age-metallicity degeneracy present in old -- intermediate age stellar 
systems. Accurate age determinations require the knowledge of the \alp element
\abus as well as \feh.

Scl also is            unusual or even unique in several respects. It appears
to be composed mainly or almost exclusively of an old, \Gal globular cluster-
aged population (e.g. Da Costa 1984, Hurley-Keller et al. 1999, 
Monkiewicz et al. 1999, Dolphin 2002). However, there are some blue stars
brighter than the turnoff (Demers \& Battinelli 1998) and it may contain neutral
H gas (Carignan et al. 1998, Bouchard et al. 2003). 
Walcher et al. (2002) find
evidence for tidal tails. Hurley-Keller et al. (1999),    Majewski et al.
(1999) and Harbeck et al. (2001)
found a gradient in the morphology of the horizontal branch, with a
much higher percentage of red HB stars in the inner regions than in the outer
regions. Harbeck et al. (2001) found Scl to have the most significant HB 
morphology gradient of any of their sample of 9 dSphs.
Majewski et al. even suggested the possibility of a bimodal \met
distribution based on their CMD. However, Tolstoy et al. (2001) found no 
indication for either a \met gradient or bimodality in their low resolution
spectroscopic
\abus for a large sample of stars.

Because of its importance, several groups independently
began high resolution studies of Scl stars. After we began our study we became 
aware of the other groups. The results of one of these groups have subsequently
been published (S03) and we have combined our results with theirs. This allows
for a substantial increase in sample size and \met coverage, allowing us to
investigate any abundance trends with much greater confidence.

An  additional aspect of a detailed analysis of Scl's red giants is
the internal evolution of the stars themselves.  
While all highly evolved, metal-poor, low-mass red giants show evidence of extra-mixing 
beyond the canonical first dredge-up (see references in, e.g., Weiss \& Charbonnel 2003), 
the degree of mixing may depend on the initial stellar metallicity. 
In both field and globular cluster stars near the red giant tip, 
the very low $^{12}$C/$^{13}$C ratio, ranging from $\sim$4 to 8,    
can be unambiguously attributed to an in situ (evolutionary) mixing mechanism. 
We can       test the universality of this phenomenon by deriving
the carbon isotopic ratios for our two most metal-rich stars. 

On the other hand, the O/Fe, Na/Fe, and Al/Fe ratios 
vary greatly from star to star in globular clusters. 
The abundance anomalies of these isotopes from higher p-burning cycles, 
which are not seen in field halo stars, seem to be mostly of primordial origin according 
to the preponderance of observational evidence 
(e.g.., Gratton et al. 2001; Grundhal et al. 2002; Yong et al. 2003). 
Massive AGB stars have been claimed to be the favorite candidates to have polluted 
the intracluster gas or the surface of cluster stars (e.g., Cottrell \& Da Costa 1981; 
Ventura et al. 2001). However Fenner et al. (2004) recently showed that the abundance 
patterns observed in a ``classical" globular cluster like NGC 6752 could not be matched 
by a model of chemical evolution incorporating self-consistently the detailed 
nucleosynthesis yields from AGB stars. 
In particular, neither the O-Na nor the Mg-Al anticorrelations could be reproduced,
in agreement with the previous findings by Denissenkov \& Herwig (2003) and 
Herwig (2004).
So the details of this primordial (or pollution) scenario still await clarification. 
By investigating the relative abundances of O, Na and Al we can see how the presence 
of these chemical variations depends on environment
and probe the connection between dSphs and globular clusters. 


Our paper   is arranged   as follows: In Section 2 we present the observations
and reductions and in 3 the details of the abundance analysis. In 4 we 
present the abundance results. 
The heavy element star we discovered is discussed
in detail in Section 5.  
In Section 6 we 
summarize our major findings. 

\section{Observations \& Reductions}

Echelle spectra of four giants in Scl were obtained on the nights of 
September 18 and 19, 2000 with the UVES instrument on the 8.2 m 
VLT UT2  telescope (Kueyen) of the European Southern Observatory. 
The target stars were selected from the study of Schweitzer et al. (1995) as
being amongst the brightest giant members (with proper motion membership 
probabilities $=99\%$) and covering the full color width near the tip of the 
giant branch. This latter should reflect to first order as complete coverage
of the full \met range as possible.
Each star was observed for a total of four hours, 
divided into one-hour exposures. The
stars were observed simultaneously through the blue and red arms of UVES using 
a dichroic beam splitter. This yields complete coverage from 
$\lambda\lambda \sim 5900 - 9600\AA$ 
in the red except for loss of a single order near the 
center where there is a gap between the two CCDs, and complete coverage from
$\lambda\lambda \sim 3700 - 5000\AA$ in the blue.
The resolving power with a $1\arcsec$ slit
is about 22,000 in the red
and 16,000 in the blue. The seeing was generally $0.5 - 1.0\arcsec$.
Spectra of hot, rapidly rotating stars were also obtained 
in order to divide out telluric absorption lines. The data were sky subtracted,
reduced to
one-dimensional wavelength calibrated spectra,
and then the individual
spectra for each star were co-added,
using the standard software packages available in IRAF.
Typical S/N ratios for the final combined spectra
are about 120 per pixel  at
6700\AA\ and 65 per pixel 
at 4500~\AA.
Radial velocities of all 4 stars, given in Table 1,
show that they are indeed members, as the mean radial velocity of Scl stars is
$108 \pm 3 km ~s^{-1}$ (Mateo 1998).  Combining our four velocities
with five velocities in S03, we find a mean of +110.0 km-s$^{-1}$, with
a dispersion (standard deviation) of 6.9 km s$^{-1}$. 
A small segment of spectrum is shown for a metal-poor star (195, with
[Fe/H]= -2.1) and for a more metal-rich star (1446, with [Fe/H]= -1.2) in Figure
1.
The difference in the line strengths is obvious; as the stars have
rather similar effective temperatures and gravities which would lead to
only very modest differences in line absorption for the same abundances,
most of the differences in the observed line strengths are caused by
abundance differences. This is the most graphic evidence for a real abundance 
spread in this dSph.

\section{Abundance Analysis}

\subsection{Stellar Effective Temperatures}

One of the fundamental parameters needed in stellar abundance
analyses is the effective temperature of the star in question.  The
procedure in this study is to base effective temperatures on two
broadband color (V--K and J--K) calibrations of T$_{\rm eff}$.  Table 1
lists the stars observed along with various apparent and absolute
magnitudes and colors and the derived radial velocities.
The star designations, V-magnitudes and (B--V) colors are taken from 
Schweitzer et al. (1995). 
The Two-Micron All Sky Survey (2MASS) database, accessed via
``http://irsa.ipac.caltech.edu'', is the source of the K-magnitudes
and (J--K) colors.  The 2MASS magnitudes and colors have
been transformed to the system defined by Bessell \& Brett (1988),
as we use T$_{\rm eff}$-calibrations from Bessell, Castelli \& 
Plez (1998), who use the color system defined in Bessell \&
Brett (1988).  The 2MASS corrections are those defined by Carpenter (2001) 
in his Appendix A and are fairly small: a constant 0.04 magnitude
offset in K, and a small color term that is about 0.03 magnitude
in (J--K).  The V-magnitudes used to compute (V--K)
are corrected for reddening based on the (B--V) color excess of
$0.02 \pm0.02$ (Mateo 1998) 
and A$_{\rm V}$= 3.3E(B--V).  No absorption correction
is applied to K, since the overall reddening to Scl is quite small 
and $A_{\rm K} = 0.1 A_{\rm V}$.
In addition, in Table 1 are the absolute K-magnitudes (for a true distance
modulus of $19.54 \pm 0.08$ - Mateo 1998), 
along with K-band bolometric corrections from
Bessell et al. (1998) and the subsequent absolute bolometric magnitudes and
luminosities.

Effective temperatures are derived for the program stars using the
(V--K) and (J--K) colors from Table 1, along with calibrations
discussed and defined in Bessell et al. (1998).  These authors point
out that (V--K) versus T$_{\rm eff}$ has almost no sensitivity to
metallicity (hereafter taken as [Fe/H]) so long as T$_{\rm eff}$ is
greater than $\sim$4000K, while (J--K) has only a small
dependency; their near independence from   [Fe/H] is why these two
colors are used.  Table 2 lists effective temperatures for each
star as defined by (V--K) and (J--K) colors.  With just the four
program stars in question, there is no systematic offset between
the two sets of T$_{\rm eff}$'s, and their differences scatter
around   100K with no systematic trend.
For the final T$_{\rm eff}$ to be used in the abundance
analyses, we adopt the average of the two values and round this to the
nearest 25K in defining the model atmosphere effective temperature
(shown in column 4 of Table 2). 
Other color -- effective temperature relations are available in the
literature and we compare one other source (McWilliam 1990) to the
calibrations used here.
Using McWilliam's (V--K) calibration and comparing it to the
values of colors and T$_{\rm eff}$'s in Tables 1 and 2, the mean difference
(in the sense of Bessell et al. (1998) minus McWilliam) and standard
deviation is found to be +7$\pm$19K: very good agreement.  Using
(J--K), the mean and standard deviation are +90$\pm$51K.  
This comparison suggests that these various temperature scales for the types
of red giants analyzed here are in good agreement with differences of 
less than about 100K. 

\subsection{Surface Gravities, Microturbulent Velocities, \& Iron Abundances}

With an effective temperature scale defined by broadband colors, the
remaining global stellar parameters of surface gravity (parameterized
as log g), microturbulent velocity ($\xi$), and overall metallicity
(defined by the Fe abundance) are set by the Fe I and Fe II lines.
The spectroscopic analysis of the Fe lines, as well as all of the
other elements to be discussed, uses a recent version of the LTE 
spectrum synthesis code MOOG, first described by Sneden (1973).  The
model atmospheres adopted are those generated from a version of
the MARCS code as discussed by Gustafsson et al. (1975).  The
combination of LTE analysis from MOOG using MARCS model atmospheres
has been used extensively in abundance analyses of K-giants, and
yields accurate chemical abundances
of many species.  Recent examples of similar analyses of red giants 
includes Ivans et al. (1999, 2001), Ramirez et al. (2001), and S01 and S03.

The microturbulent velocity, at a given log g, is found by forcing
all Fe I lines to yield the same iron abundance, i.e., with no
significant slope of A(Fe)\footnote{ A(Fe)= log[N(Fe)/N(H)]+ 12.0.}
versus reduced equivalent width 
(log (W/$\lambda$).  Once the microturbulent velocity is defined,
the surface gravity is tested for consistency by comparing the
Fe I and Fe II abundances, with ionization equilibrium demanding 
both neutral and singly ionized Fe to yield the same abundances. 
The now determined stellar parameters (T$_{\rm eff}$, log g, and
$\xi$) along with the Fe I and Fe II lines are then used to derive
the overall metallicity  
of the stellar atmosphere.  The entire process of defining log g,
$\xi$, and [Fe/H] is iterated until a consistent set of stellar
and model atmosphere parameters is found, and this atmosphere is finally
used in the derivation of the other elemental abundances.

When enforcing the ionization equilibrium of Fe I and Fe II to derive
surface gravity in metal-poor red giants, some care should be taken
to investigate the possible influence of over-ionization on Fe I,
as discussed by Thevenin \& Idiart (1999).  Data from three recent
analyses of globular cluster giants can be used to partially address
this question: Ivans et al. (1999) for M4, Ramirez et al. (2001) for
M71, and Ivans et al. (2001) for M5.  In all these studies, 
``evolutionary'' gravities are calculated from stellar model tracks
(with known masses) coupled to the derived red-giant effective
temperatures and luminosities.  Iron abundances are then computed and
can be compared between Fe I and Fe II.  From the three studies 
noted above, no significant effect is seen over the metallicity range
from [Fe/H]= -0.84 (for M71 with $\Delta$(Fe I -- Fe II)= +0.14$\pm$0.17),
to [Fe/H]= -1.15 (for M4 with $\Delta$(Fe I -- Fe II)= -0.01$\pm$0.09),
and [Fe/H]= -1.21 (for M5 with $\Delta$(Fe I -- Fe II)= -0.13$\pm$0.07).
Based on the above observational constraints, overionization may affect 
derived gravities by $\sim$ 0.1 dex, and possibly lead to small over/underestimates
of the iron abundance by $\sim$ 0.1 dex.  Such a possibility will not, however,
have a significant effect on the derived abundances and abundance trends
to be discussed here.

The derived effective temperatures, gravities, microturbulent velocities,
and iron abundances are listed in Table 2.  The iron abundances are
derived using the accurate sets of gf-values from Martin, Fuhr, \&
Wiese (1988), Bard, Kock, \& Kock (1991), Holweger et al. (1991), 
and O'Brian et al. (1991).  It is worth noting that the absolute
accuracy of these Fe gf-values is now at the few percent level, and
an analysis of solar Fe I and Fe II lines yields photospheric iron
abundances with a scatter of $\sim$ 0.05 dex and essentially perfect
agreement with the meteoritic abundance (A(Fe)= 7.50 - 
Grevesse \& Sauval (1999)). 
Table 3 gives relevant parameters for all of our measured Fe lines.

We will later combine our abundances to those derived by S03 for five
other Sculptor red giants in order to create a larger database.  
Abundances will be given as values of [X/Fe] so we must check for any
offsets in the two abundance scales caused by either gf-values or adopted
solar abundances.  In the case here for iron, we note first that S03
used A(Fe)= 7.52 for the Sun, whereas we adopt 7.50; to strictly
compare our respective Fe abundances, we should add +0.02 dex to the
S03 values to bring them onto our scale.  In addition, however, a
comparison of gf values for
Fe lines reveals that for 15 lines in common, there is a
mean offset of +0.04 dex in log gf (in the sense of Us -- S03).  Based on
this difference (if it is indicative of a general trend for all Fe lines), 
we should then subtract 0.04 dex from the S03 [Fe/H] values to bring them
onto our gf-scale.  The net result of considering the adopted Solar Fe 
and gf-value scales would be to shift by -0.02 dex the S03 values of [Fe/H]
to bring them into agreement with our scale.  This is such a small
offset, well within the uncertainties of the gf-values themselves and even
the Solar Fe abundance, that we consider such a difference to be
insignificant and apply no corrections to the S03 values of [Fe/H]: both
studies are effectively on the same abundance scale.

\subsection{Elements Other Than Iron} 

Of the spectral species studied
here in common with Smith et al. (2000), we have adopted the gf-values
from that study, and their sources are discussed in detail in that
paper.  These elements are O I, Na I, Mg I, Al I, Si I, Ca I,
Sc II, Ti I, Ti II, Fe I, Fe II, Ni I, Y I, Y II, Zr I, Zr II,
Ba II, La II, and Eu II.  Smith et al. (2000) analyzed
both the Sun and Arcturus using this linelist and gf-values and derived
expected abundances, indicating that these gf-values can be used to
derive accurate abundances.  In addition, we have added the species 
Mn I and Zn I in this study, with the Mn I gf-values taken from
Prochaska \& McWilliam (2000).  Neutral zinc gf-values were adjusted to
yield a Solar abundance of A(Zn)=4.60 using a 1-d MARCS model; this resulted
in log gf= -0.44 for the 4722\AA\ line (in excellent agreement with the
value of -0.39 from S03), and log gf= -0.24 for the 4810\AA\ line (again
in excellent agreement with S03 who used -0.17).  A comparison of the
other elemental gf-values with the same lines used by S03 finds differences
of less than 0.05 dex in log gf for all cases except Al I.
In the case of differences having less than 0.05 dex, no offsets will be
applied to the S03 abundances when adding them to our dataset.  For
Al I the offset is +0.26 dex (in the sense of Us -- S03),
this difference was       applied in order to combine
our Sculptor aluminium abundances with those from S03. 

There are also a few other points concerning gf-values that should
be noted.  Two Ca I lines have differing log-gf values greater than
0.05 dex; the 6439.08\AA\ gf-value used here is 0.08 dex larger than
in Shetrone et al. (2003), but this is only marginally larger than
0.05 dex.  The gf-value for the 6161.30\AA\ line (Smith et al. 2000)
is 0.24 dex larger than that adopted by Shetrone et al.; however the
particular value used here, along with a MARCS solar model,
yields a solar calcium
abundance of A(Ca)=6.25.  This is acceptably close to the recommended
value of A(Ca)=6.34 from Lodders (2003).  As calcium is represented
by 8 lines in this study, and 9 lines from Shetrone et al., with the
other gf-values in close agreement, the differences discussed above
will not affect significantly the average Ca abundances.

The Mg I lines used here at 8717.83\AA\ and 8736.04\AA\ were not in
Smith et al. (2000) and the gf-values here are taken from the
Kurucz (1991) compilation.  Analysis of these lines in the solar
flux spectrum results in respective magnesium abundances of A(Mg)=7.51
and 7.54, close to the Lodders (2003) recommended value of 7.55.

There are no lines in common between this study and that of Shetrone
et al. (2003) for Ti I, Ti II, Y I, and Y II, but later inspections
of Figure 6 (for Ti) and Figure 8 (for Y) will find no significant
differences in the respective behaviors of Ti and Y with metallicity
between the two studies.

Hyperfine splitting (hfs) was included
for the species Sc II (with the hfs data taken
from Prochaska \& McWilliam 2000), Mn I (hfs data also taken from Prochaska
\& McWilliam 2000), Y I (with hfs data taken from Biehl 1976),
La II (with hfs data taken from Lawler, Bonvallet, \& Sneden 2001), 
and Eu II (with hfs data taken from Biehl 1976).  For those species with
multiple isotopes, solar isotopic ratios were assumed. 

Table 4 provides the relevant parameters for all of our measured 
non-Fe lines, as well as their equivalent widths in the Sculptor red
giants.  Although we present the equivalent widths, it must be noted
that all abundances were derived via spectrum synthesis.  The final
abundances are given in
Table 5, in the form of [X/H] values, as well as the adopted solar
abundances in the form of A(X). 

As discussed in Section 3.1, various comparisons between (V-K) and
(J-K) temperature calibrations can reveal systematic differences
of up to 90K, with a scatter of 50K, thus an expected uncertainty
($\sim$1$\sigma$) of about 100K for T$_{\rm eff}$ in red giants
is a reasonable value.  In addition, the 1$\sigma$ scatter set by the
Fe I lines in defining a mean iron abundance is about 0.15 to
0.20 dex; this scatter is carried into the determination of the surface gravity
from the Fe II lines, and leads to an uncertainty here of about
0.3 dex in log g.  Finally, the microturbulence is defined by using
the Fe I lines (with the criterion of no trend in Fe abundance with
reduced equivalent width) and the minimum scatter in the Fe I abundances
leads to an uncertainty in $\xi$ of about 0.3 km s$^{-1}$.  The
uncertainties of $\pm$100k in T$_{\rm eff}$, $\pm$0.3 dex in log g,
and $\pm$0.3 km s$^{-1}$ represent approximate 1$\sigma$ values for
these fundamental stellar parameters.

All of these elemental species present different sensitivities in their
derived abundances to the primary stellar parameters of T$_{\rm eff}$,
log g, and $\xi$.  Table 6 quantifies these sensitivities for star 770,
which is near the middle of our sample in terms of effective temperature,
gravity, and micorturbulence; the other stars will exhibit very similar
sensitivities to changes in stellar parmaeters.  The differences for
each species are tabulated for a change of +100K in T$_{\rm eff}$,
+0.3 in log g, and +0.3 km-s$^{-1}$ in $\xi$, with the final column
showing the quadratic sum of these uncertainties.  This final value is
a fair estimate of the uncertainty in the derived abundances caused by
realisitic uncertainties in defining the fundamental stellar parameters.

\section{Abundance Results}

\subsection{Iron/Hydrogen}

We first                                   combine our data with
that of S03 obtained with the same telescope and 
instrument. 
This gives us a total of 9 stars in Scl, the second largest sample of high 
resolution abundances yet obtained for a dSph, surpassed only by
several Sagittarius studies,
with a range in [Fe/H] from --2.10 to --0.97.  Note that this 
substantially extends the \met range of S03's sample, which covered --1.95 to
--1.2. We also cover most of the known \met range in this galaxy, based both
on the extreme colors of our sample in observed CMDs and the large sample of
Ca-triplet abundances derived by Tolstoy et al. (2001) for 37 stars. Other
than 2 stars with \feh of
--2.2 to --2.3, their next most \mp star has \feh$=-1.94,
$ and at the \mr end their distribution stops at --1 except for a single star
with --0.8.
In particular, our combined sample 
allows us to see abundance trends with \met
more clearly. Also note that our spectra are of significantly higher S/N
than those of S03's Scl observations.

Our mean \feh$=-1.57\pm 0.12$ (standard error of the 
mean). This compares very well with the mean of $-1.5\pm 0.3$ found by 
Tolstoy et al. (2001) and the mean of --1.5 found by Dolphin (2002).
We do not find any indication of a \met gradient, although our 
sample is small and only covers a limited radial
range, from 0.1 -- 1.1 core radii.
Tolstoy et al. (2001) have a much larger sample and radial extent and found no
gradient.

\subsection{Sodium and Oxygen}

Sodium is produced in the Ne-Na cycle in which protons
are captured by the Ne isotopes accompanied by the necessary $\beta$-decays.
The stellar environment for the Ne-Na cycle is uncertain but it can occur
at relatively low temperatures, near 30 million K, in 
evolved
stable stars. In
addition Na can be produced by carbon burning which requires temperatures
that can be realized only in advanced stages of stellar evolution and in massive
stars. In our nine
Scl stars the mean [Na/Fe] ratio is $-0.47\pm 0.06$,
with no trend with \met. This ratio differs from recent 
analyses of  globular clusters in which an excess of Na is usually
seen. In M15 and
M92 Sneden, Pilachowski and Kraft (2000) find [Na/Fe]=+0.2 on average. In M5
Ivans et al (2001) find [Na/Fe]$\sim$ -0.2 in oxygen-rich stars, rising to +0.4 
in oxygen-poor stars.
In M4 Ivans et al. (1999) 
find [Na/Fe]$\sim$ 0.0 for stars that do not show evidence of 
a deficiency of oxygen. 
Clearly the Na deficiency seen
in Scl differs from the situation in the globulars. 
In the halo field, Fulbright (2002) found that [Na/Fe] scattered around
0.0 all the way down to [Fe/H]= --4.0 with some stars showing [Na/Fe] as
large as +0.5 between [Fe/H]= --2.0 to --3.5 and some scattering down to
[Na/Fe] = --0.5 between [Fe/H]= --1.5 to --2.5.
Indeed the stars in his ``high velocity" bin 
have [Na/Fe] very similar to our sample.
Even if a correction of +0.2
is added to our [Na/Fe] values to compensate for potential non-LTE effects
(e.g. Tautvaisiene et al. 2004), our Na abundances are still
$\sim 0.3$  dex  lower than 
typical     Galactic stars of similar metallicity.

Our oxygen abundances were
derived from the  single 6300\AA\ line of [OI], while S03 used both this line
(mainly) as well as the 6363\AA\ line when available (2 stars).
Although generally weak and only a single line, we feel that our O abundances
are well-determined via the spectrum synthesis technique.  A comparison of
observed and synthetic spectra covering the [O I] line in Star 1446 is
shown in Figure 2 to illustrate the quality of the spectra and corresponding
synthetic matches.  
In Figure 3 we show the correlation of [O/Fe] with [Fe/H] for both the
Sculptor giants (8 stars sampled, with 4 from this study and 4 from S03),
and a sample of Galactic field stars from a number of studies (noted in
the figure caption).  The Galactic studies shown in Figure 3 include those
that rely either on the [O I] 6300\AA ~line, or the infrared 
vibration-rotation lines from OH. 
At the metal-poor
end, the [O/Fe] value is similar to, albeit a bit lower  in the mean than,
that which is seen in metal-poor globular
clusters and the halo field (McWilliam 1997).  However, the ratio of O/Fe
decreases steadily and rapidly as \feh increases above --1.5,
reaching [O/Fe]$\sim$ 0.0
at [Fe/H]= --1.2, and the most \mr star at \feh $=-1$
has a very low [O/Fe] abundance of --0.3.
This is distinctly different from the globulars and the halo and disk
field where [O/Fe] only begins to fall dramatically
 at [Fe/H]=--1 and does not reach zero
until [Fe/H] is near zero. The O abundance of Star 982 is some 0.6 dex lower
than the mean for the \Glx  at this \met. Note that this trend is seen
much more clearly here than in S03 due to the extended \met range, or for
that matter from our data alone, which suggest a strong monotonic decline.

As noted in the Introduction, it is very important to check for the existence
of any Na-O (anti)correlation in order to probe the connection between
dSphs, the Galactic halo and globular clusters.
In Figure 4 we present [Na/Fe] vs. [O/Fe] for the Sculptor red giants 
along with the data for Galactic field stars. There is no trend in Sculptor, 
although note that even the low-O star 982, the heavy element star 
(see Section 5),    shows [Na/Fe]= -0.2.
Where the anti-correlation is seen, it has often been ascribed to deep mixing
within the star which has carried material to the stellar surface that had
been processed by the ON, NeNa  and MgAl cycles 
(see Dennisenkov and Weiss 2001 and references therein). 
However recent observational evidence and theoretical developments 
(see Charbonnel \& Palacios 2003 and references therein) now point toward 
an alternative explanation. 
Namely, these anomalies may be due to stochastic enrichment by an
earlier generation of stars 
followed by scattered incorporation into some, but not all, of the stars 
now seen in globular clusters (Cottrell \& Da Costa 1981; D'Antona et al. 1983; 
Jehin et al. 1998; Parmentier et al. 1999). 
However the major source of pollution for GCs still remains to be determined. 
Indeed, based on current theories of intermediate mass stellar evolution 
and nucleosynthesis, it now appears that massive AGBs, which seemed to be the 
most plausible candidates for this pollution, are not responsible for
the observed globular cluster abundance anomalies (Denissenkov \& Herwig 2003; 
Herwig 2004; Fenner et al. 2004). 
Since we see neither the O/Na nor the O/Al anti-correlation in the
Sculptor stars it appears that neither of the above scenarios (in situ extra-mixing 
of NeNa and MgAl cycle material or pollution) that induced the anomalous chemical 
patterns in globular clusters has occurred in this dwarf galaxy.
The anti-correlation is never seen in field stars of the Galactic halo.
These facts tend to associate Sculptor with field stars rather than with
globular cluster stars. We note, however, that McWilliam et al. (2003)
have found one star in Sgr with a mild excess of [Na/O].

The other light odd element is Al for which our fragmentary data 
(we only derive Al
abundances for the two most \mr stars) shows a deficiency
similar to that of Na, while in globulars the [Al/Fe] value varies, reaching
as much as +0.6 in M4 in stars that show no evidence for the depletion
of oxygen (Ivans et al. 1999).

\subsection {The $^{12}C/^{13}C$ Ratio}

While the Na/O ratios in Scl do not provide evidence for deep mixing, 
the ratio of $^{12}$C/$^{13}$C in two stars shows that moderately deep mixing
has indeed occurred. For our two most metal-rich stars it was possible to
derive the carbon isotope ratios from the 8005\AA\ $^{13}$CN feature. 
For star 1446
the derived ratio is 3 with a substantial uncertainty. For the heavy element
star, 982, the $^{13}$CN feature is strong enough for an estimate of the
uncertainty and we find a ratio of $3.5\pm1$. We show the fit for star 982 in
Figure 5.  These low ratios of $^{12}$C/$^{13}$C are almost exactly the CNO
equilibrium ratio and show that the vast majority of the material presently
in the stellar atmosphere has been subject to proton capture at temperatures
of at least 10$^{7}$K. 
Low $^{12}$C/$^{13}$C ratios
are the general rule for stars near the red giant tip 
in the field (Charbonnel, Brown \& Wallerstein 1998; Gratton et al. 2000), 
and in galactic open (Gilroy 1989; Gilroy \& Brown 1991) and globular clusters 
(see Charbonnel \& do Nascimento 1998 for early references; 
Smith et al. 2000; Shetrone 2003). 
The same is true for RGB stars in the LMC (Smith et al. 2002) and in the SMC 
(Hill et al. 1997). 
Our results thus confirm the universality of an extra-mixing mechanism 
which transports matter between the outer layers of the hydrogen burning 
shell and 
the convective envelope in all low-mass red giants, and which is still not 
part of the standard evolution theory. This extra-mixing is independent
of the stellar environment and metallicity, although it has been shown
to depend on the initial stellar mass. 
 
\subsection{Alpha Elements}

For the purposes of this Section, we will include here as
alpha elements Mg, Si, Ca and Ti. See S03 for a more thorough discussion of
the various nucleosynthetic origins for the alpha elements.
Silicon scatters about a
mean near solar while
magnesium, calcium, and titanium show
interesting and similar trends albeit at different significance levels,
as shown in Figure 6. 
All three elements show trends generally similar to that of O, with
solar or enhanced and
relatively constant values for the stars with \met $<-1.5$ that are generally
less than those of normal
Galactic field stars at similar \mets, but the more \mr 
stars have significantly lower \abus that are generally much lower than
those of Galactic stars of the same \met. The \abu deficit with respect to 
Galactic field stars is in fact true for Si at all \mets as well.
In the case of Mg, the lowest \met star has the most
enhanced \abu and the trend of decreasing \abu over the full \met range
is the most pronounced.
The decrease in \abu for the most \mr stars is smallest for Ti and indeed
the difference between these and the more \mp stars is not very significant.
[Ti/Fe]
is nearly constant at about solar with only a small decrease             
to --0.2 for the three stars near --1. 

This observed behavior in the Scl stars
is different from the trends in the globulars and
the halo where [Ti/Fe] is usually near +0.3 over the range of [Fe/H] from
-2 to -1; though some metal-poor globulars show nearly the solar ratios
(Lee and Carney, 2002).
In fact, models of Type II supernovae
usually place Ti as an iron-peak element rather than associate it with the
alpha-elements (Arnett 1996). 
It appears that in Scl Ti follows SNII models better
than do most of the globulars and field stars of low metallicty.
But models comparing the relative yields of SNII and Ia (e.g. Lee and Carney
2002, Fig. 9, based on Woosley and Weaver 1995 models) 
show that Ti is {\it the} element most produced in SNII's compared to SNIa's. 
Shetrone (2004) suggested that the ``light"
(O and Mg) and ``heavy" (Ca and Ti) \alp elements may show differences in 
their behavior due to potential differences in their nucleosynthetic origins.
We find that                                       the mean enhancements of
Mg and Ca in Scl (0.08 and 0.11 dex) are very similar but O (0.22 dex) is
significantly enhanced with respect to Ti (-0.08 dex) and thus there is no
clear trend.

\subsection{The Iron Peak} 

All of the elements from Sc to Zn are usually ascribed to the Fe-peak though
a few of the odd elements may be enhanced by small neutron-capture processes. 
The relative abundances of Sc, Mn, Cu, and Zn are shown in Figure 7.
In Galactic metal-poor stars Sc follows Fe very uniformly 
(McWilliam 1997). In Scl,
between [Fe/H]=-2.1 and -1.5, Sc
follows Fe just as in the Galaxy, although the mean
[Sc/Fe] is slightly less than in the
Galaxy. But the higher \met Scl stars deviate
significantly from their Galactic counterparts: [Sc/Fe] begins to decrease at
a \met of --1.5 and drops rapidly to about --0.5 by
[Fe/H]=-1.0, very reminiscent of the general \alp behavior.
We know of no other object in which this phenomenon occurs.    
For Cr we must rely on the data of S03 whose Cr/Fe ratios
for 5 stars lie close to -0.15 dex which is very similar to field stars in
the halo (McWilliam 1997, Fig. 12a). The mean [Mn/Fe] ratio is near -0.4
which is similar to that in halo field stars of
the same metallicity interval.  However the [Mn/Fe] values for the largest 
and smallest deficiencies of Fe suggest a small downward trend in 
[Mn/Fe] with [Fe/H].  While the reality of this trend, which only depends on
the two endpoints,
is uncertain, especially
when the error bars are taken seriously, the trend appears to be
real.  If so, this trend is contrary to the rise of Mn with Fe seen in the
Galactic halo, thick disk and bulge (e.g. McWilliam et al. 2003).
For Co/Fe the 4 
stars analysed by S03 
also track iron as do the halo field stars with [Fe/H] $>-2.0$. 
The same
holds for the Ni/Fe ratio.
For Cu the story is very different.  In the 5 stars for which we
have derived copper abundances, with [Fe/H] between -1.2 and -1.8,
the values of [Cu/Fe] are very near -1.0 with a hint of a downward trend with 
\feh, significantly
different from the field halo stars and stars in $\omega$ Cen,
where it is close to -0.5 in the same
interval of [Fe/H] (McWilliam 1997, Cunha et al. 2002). 
The field stars show a significant slope, reaching -0.8 dex in [Cu/Fe]
versus [Fe/H] at [Fe/H]=-2.5.  Mishenina et al. (2002) find a steady 
decrease in [Cu/Fe] with decreasing [Fe/H].  However, over the interval 
of [Fe/H]= --1.2 to --1.8, their data are consistent with [Cu/Fe]=--0.5.  
Finally for Zn there appears to be no mean trend away from
[Zn/Fe]$\sim$ 0.0, as seen in the halo field stars, 
but it appears that there may
be a bifurcation, with one group of Scl
stars with [Zn/Fe]$\sim+0.15$ and another
with [Zn/Fe]$\sim -0.35$.

\subsection{Heavy Elements}

Elements heavier than Zn are known to be produced by neutron capture. The
two neutron capture phenomena are the slow (s-process) capture sequence in
which the time scale for beta-decay is shorter than the time scale between
neutron captures and the rapid (r-process) capture sequence in which a flood
of neutrons (or a
burst of nuclear reactions whose results mimic neutron captures)
drives the nuclei to extremely neutron-heavy isotopes which
finally decay to the valley of stability. The readily observable s-process
species may be divided into the light (ls) group which consists of Rb,
Sr, Y, and Zr; and the heavy (hs) group which consists of Ba, La and the
light rare earths.
The only observable r-process element in our stars is Eu. The 
heavier r-process species have lines that are too weak to be 
measurable on our spectra. Starting with the ls elements,
either Y, Zr, or both are available in 8 stars
of the combined data. Ignoring for the moment the most \mr star (982), 
the mean ls \abu (using the average of Y and Zr when available) is --0.26
with a large $1\sigma$ scatter of 0.4 dex and no \met trend. 
This value is very low compared to Galactic field stars (e.g. Fulbright 2002).
As shown in Figure 8, star 982 is unique, standing far above all the other
stars with [Y/Fe]= 1.1. This is a heavy element star! 
We discuss this star in detail in Section 5.  
The [hs/Fe] data (taking the mean of Ba and La) scatter around a value of
0.1 with no trend, as shown in Figure 9
(omitting the heavy element star 982 which again is remarkable, as seen
from this figure). 
With two     exceptions, the Scl stars mimic their \Gal counterparts in [hs/Fe].
[Eu/Fe] (Figure 10) appears to 
decrease from near +0.7 at [Fe/H]$\sim-2$ to $\sim$ solar at [Fe/H]=--1.2,
although the trend again is mainly determined by only two stars.
Once again we omit the heavy element star 982 in which there is a huge
excess of Eu. 
With the exception of these three stars, the remainder are in  good agreement
with the mean for similar \met \Gal stars.

Perhaps the best discriminator of the relative importance of s-process to 
r-process enhancement is the behavior of [Ba/Eu] as a function of \feh. In 
Figure 11 we present our Scl results. The long dashed line below
represents the pure r-process \abu ratio from Arlandini et al. (1999), while 
the long dashed line above represents the pure s-process \abu ratio.
In the Galaxy, this ratio is $\approx$ constant at $\sim -0.4$ throughout the halo
\met regime but begins to rise at \feh $\sim -1$ and reaches a solar value 
near \feh=--0.4, 
significantly above the r-process line but well
below the s-process value. Scl has a unique behavior in this diagram. The
most \mp stars follow the \Glx. However, the stars more \mr than \feh=-1.5
follow an upward trend reaching to [Ba/Eu]=+0.5 at \feh=--1, some 0.8 dex 
above the \Gal value. Note that the heavy element star 982 does not stand out
in this diagram as being unusual but only extends the trend begun  by its
nearest companions in \met.

The rise in [Ba/Eu] for Scl from Figure 11 occurs near    [Fe/H]= -1.5,
while in field stars no
such rise is seen until [Fe/H]$>-1$.   In $\omega$ Cen, however, there is a
much steeper increase between [Fe/H]= -2.0 and -1.4, as shown in 
Figure 11, plotted as the filled squares.  Other s-process species show
a similar rise in $\omega$ Cen (e.g.  Vanture,
Wallerstein, \& Brown 1994, Norris \& Da Costa 1995, Smith et al. 2000). 
In the Galaxy,         note that the metal-poor stars have
an $\approx$ r-process ratio, which gradually increases towards the s-process
value as [Fe/H] increases; this reflects the increasingly important
contribution to the heavy elements from AGB stars as the Galaxy evolves
chemically.  $\omega$ Cen is known to be heavily influenced in its
chemical evolution by AGB stars and this is shown by the rapid increase
(in terms of an increase in [Fe/H]) to a pure s-process ratio in Ba/Eu
as Fe increases. The s/r-process
chemical evolution in Scl appears to have been intermediate
to that experienced by the Milky Way and $\omega$ Cen.
However, note that Johnson and Bolte (2001) found that the
interpretation of Ba in this metallicity
range is complicated.  Galactic Ba abundances from the 4554\AA\ line
can show a rise in Ba/Eu occuring at very low metallicities (around --2.0),
which is very unlikely to be due to s-process enrichments since they
are not matched by La/Eu enhancement.   This line is not used in our 
analysis nor that of S03.
Venn et al. (2004) showed that
this Ba/Eu early rise is also seen in other dSph stars.
 

\section{The Heavy-element Star 982}

Figure 12 compares a small portion of one of the orders from the 
spectrum of stars 1446 (top) and 982 (bottom). Note that these stars have 
similar effective temperatures and Fe abundances, as evidenced by 
the similarity
in the strengths of the FeI lines.  However, note the immense strength
of lines due to the s-process elements zirconium and barium, 
as represented by Zr I and Ba II,
in the spectrum of star 982. 
The unusually large ratio of all species from Y to Eu relative to Fe mark 
Sc982 as a heavy element star. 
Such stars are extremely rare in 
our halo but several have been found now in dSphs (S01, S03).
There are three types of heavy element stars.
Lloyd Evans (1983) suggested that the heavy element stars in $\omega$ Cen were
formed with their observed heavy-element excess. Many Galactic heavy-element
stars, referred to as ``intrinsic", have generated their own excess heavies. 
The best example of such objects
are the S stars which contain technetium.
A third type of heavy-element star has received a dose
of heavies from a now defunct companion. They are referred to as ``extrinsic",
and are identified by the fact that they are spectroscopic binaries 
with periods near a year or somewhat greater. 
We have searched for the Tc I lines in the 4238-4297 \AA ~region in Scl 
982 and have not found them. Due to the heavy blending and modest S/N
of the spectrum, this test is not definitive, but is 
indicative that we are not dealing with a recently 
self-polluted S or SC star.
We have no information on a possible variable radial velocity of 982, 
but its observed velocity falls within the spread of the other stars.

Nevertheless we can derive some useful information about the star or stars
that produced the heavy elements seen in Sc982. A useful spectroscopic
criterion is the hs/ls ratio with the ls species represented by Y and Zr and
the hs represented by Ba and La. In addition we have measured the Rb abundance 
which is sensitive to the neutron density during the neutron capture events
that added to the heavy elements. Using our observed value of 0.6 for [hs/ls]
and Figures  17 and 18 of Smith (1997), we
find a neutron exposure of tau = 1.1(mb${^-1}$) and a 
log N(n) = 8.6 (cm${^-3}$).
In addition, the ratio of 0.6 for [hs/ls] combined with the metallicity of
[Fe/H]=-1 places Sc982 among the CH stars (Vanture 1992) though it does not
appear to have the enhanced CH and C$_{2}$ shown by CH stars. Hence it is
possible that the star that produced the heavies was a CH star. 

One phenomenon seen in many red giants in globular clusters (but not in
field stars) is a deficiency of oxygen combined with an excess of
Na, as discussed above.  The enhanced Na/O ratio has been explained with proton captures 
by both $^{16}$O and the NeNa cycle, either within the observed red giant
or by AGB stars that enriched the presently observed star.  A significant
enhancement of the Na/O ratio requires a temperature near
30 x 10$^{6}$K for the required proton captures to be effective on a
reasonable timescale.  The abundances of star 982 as shown in Table 5
indicate a low             O/Fe when compared to other Sculptor
red giants: [O/Fe]= -0.30, while the mean for the other stars is 
+0.30$\pm$0.20.  The plot of [Na/Fe] versus [O/Fe] shown in Figure 4
shows no strong trend, although star 982 has the lowest [O/Fe] and
largest [Na/Fe] values. 

\section{Summary                               }

In this paper we have combined our new VLT plus UVES high resolution
abundance data for 4 stars in the Sculptor dwarf spheroidal galaxy
with similar data obtained with the same instrument
for 5 stars by Shetrone et al. (2003). This extends the range of [Fe/H] covered from
-2.1 to -0.97 and allows us to 
distinguish a number of interesting 
trends of various elemental abundances with Fe that were either not visible or
only hinted at in the more limited dataset of S03.
The most important single fact that emerges is that from oxygen to
manganese many elements show a relatively constant elemental ratio [X/Fe]
at the \mp end and then declining rapidly for \feh $>-1.5$, or a
steadily decreasing value of [X/Fe] as [Fe/H] rises
from -2 to -1.
The elements showing this behavior include O, Mg, Ca, Ti, Sc and Mn
(while the
Ba/Eu ratio increases with \met above \feh = --1.5).
It is  unique  to see the same pattern for all of these
elements.  In
particular the ratios of [O,Mg,Ca/Fe] near [Fe/H] = -2 are similar to but
slightly less than their
values in the Galactic halo at similar \met,
but their decline to $\sim $  
solar or less
near [Fe/H] = -1.0 rather than at [Fe/H] = 0.0 is unique. [Sc/Fe] and [Ti/Fe]
are near solar at [Fe/H] = -2 and become substantially negative at [Fe/H] = -1. 

One of our prime motivations for undertaking this study was to further test
the hypothesis that the halo of our \Glx may have been accreted from dSph-like
objects such as Scl, as first proposed by Searle and Zinn (1978).
Our derived composition of the Scl stars does
not support the suggestion that the halo of our Galaxy was formed from stars 
such as those now seen in Scl. This point has already been made by S01,
F02, S03 and Tolstoy et al. (2003). We find that Scl stars are significantly underabundant in
[\alp/Fe] at all \mets
with respect to 
typical Galactic field stars. 
AGB stars in Scl were more important in the chemical 
evolution of Scl than in the \Glx
in causing the high s-process/r-process ratios ocurring in the most \mr stars.
Finally, we find a heavy element star, with 
very strong enhancement of s-process elements.


We thank Paranal Observatory for the excellent support received during our
observing run, especially from T. Szeifert and G. Marconi. 
We would like to gratefully acknowledge M. Shetrone, E. Tolstoy, V. Hill, K.
Venn, A. Kaufer and F. Primas for allowing us to access their results prior to publication.
T. Richtler made valuable comments on an earlier draft.
Special thanks to the referee, K. Venn, for a very helpful, thorough and
constructive job which significantly improved this paper.
This work is supported in part by the National Science Foundation through
AST99-87374  and AST03-07534 (VVS).  The contributions by GG and GW 
were supported by the 
Kennilworth Fund of the New York Community Trust. 
D.G. gratefully acknowledges support from the Chilean
{\sl Centro de Astrof\'\i sica} FONDAP No. 15010003. D.G. also kindly
recognizes the warm support and constant dedication of M.E.B.
C.C. thanks the French Programme National Galaxies and the Swiss National 
Science Foundation for financial support.
\clearpage



\begin{figure}
\plotone{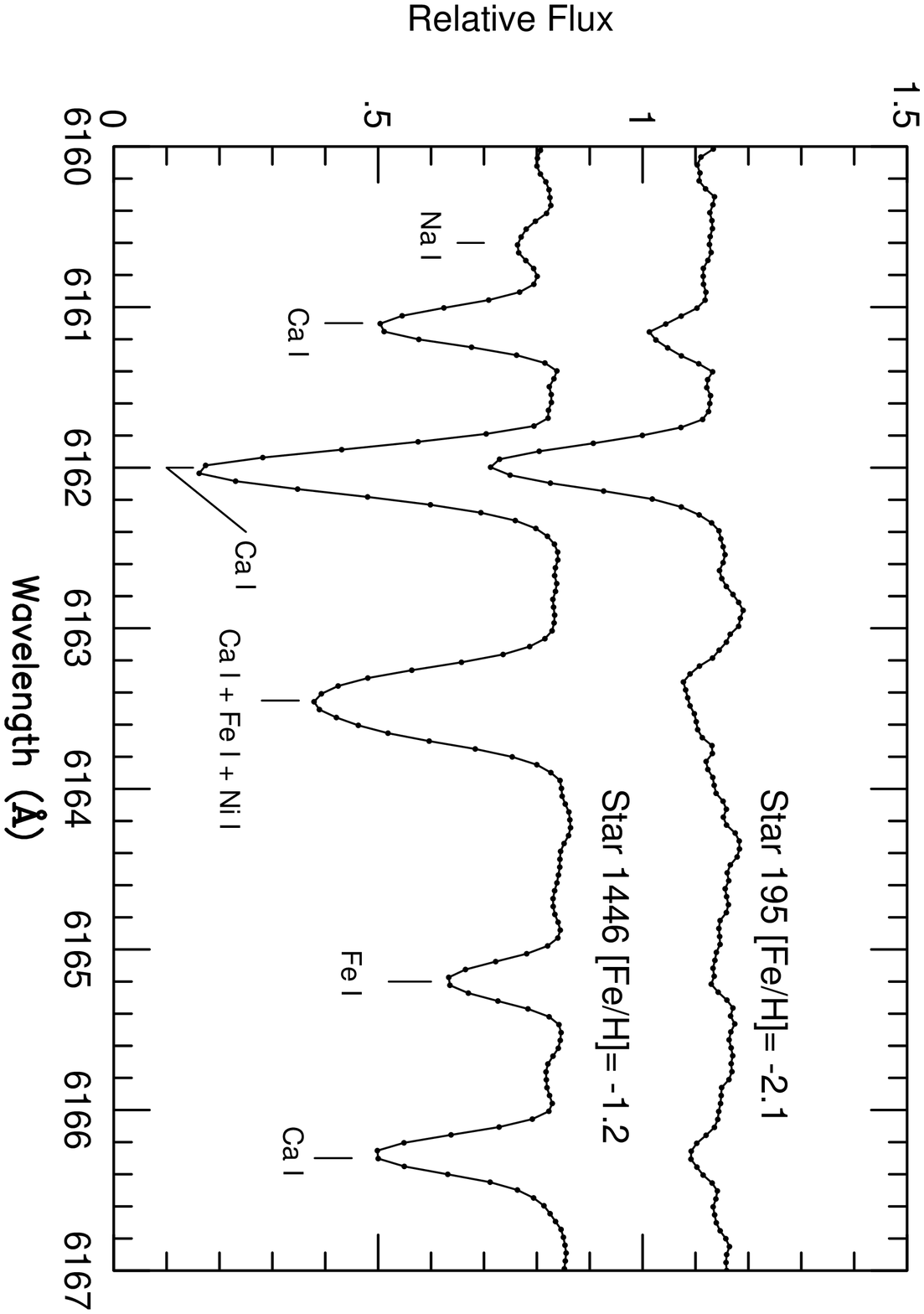}
\figcaption[fig1.ps]{
Sample spectra of two Sculptor red giants which have different overall
metallicities.  The absorption lines differ in strength due primarily
to the metallicities, with rather small differences in effective
temperature and gravity playing almost no role. 
}
\label{fig1}
\end{figure}

   \begin{figure}
   \centering
    \includegraphics[width=16cm]{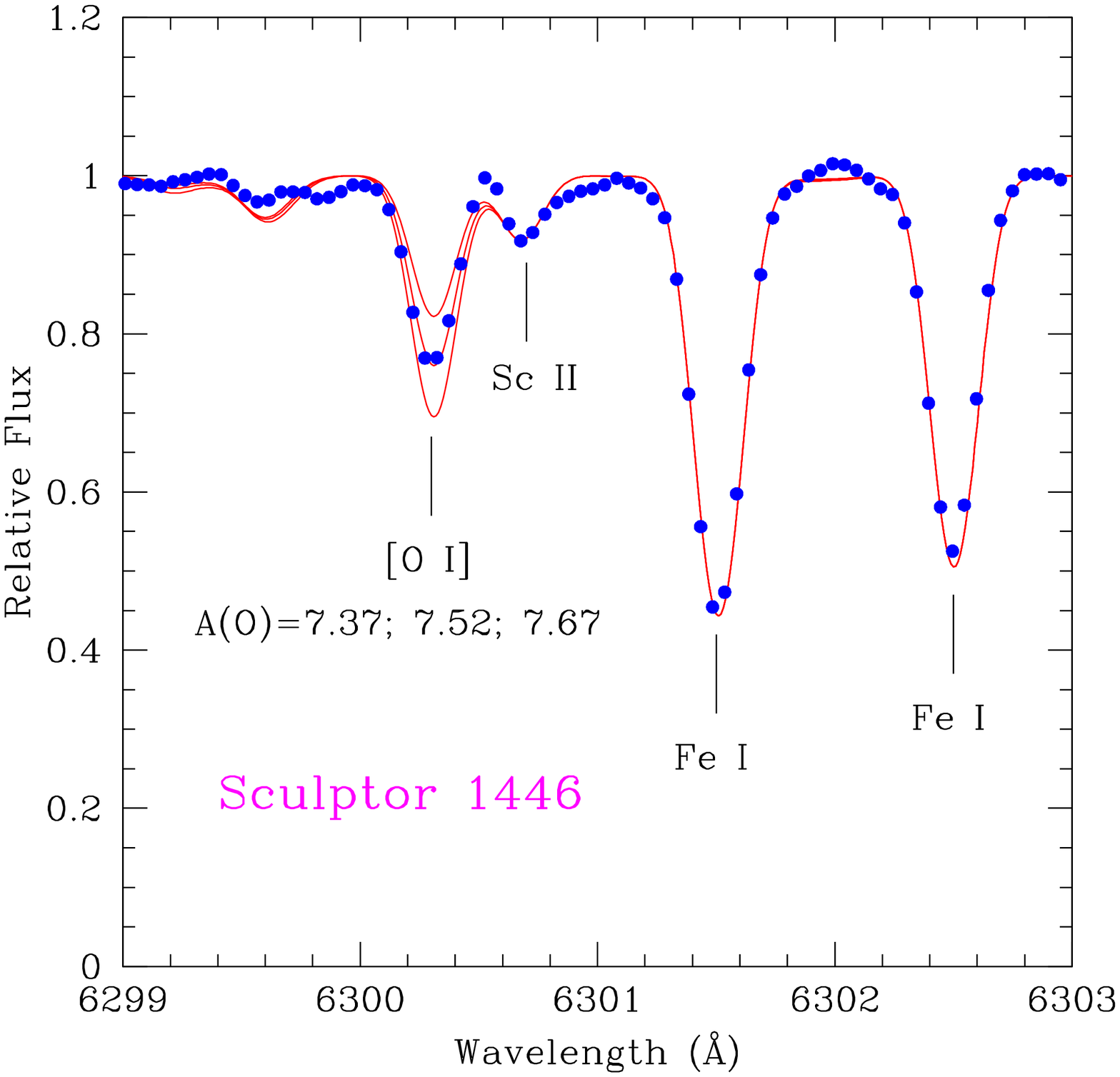}
      \caption{Comparison of the     observed spectrum for star 1446 and three
synthetic spectra computed with different oxygen abundances, spanning a
factor of 2 (0.3 dex), with the fit being done to the [O I] 6300\AA\
line.  These spectra illustrate both the quality of the observed spectrum
and how well the synthetic spectra compare to the real one.}
         \label{fig2}
   \end{figure}

\begin{figure}
   \centering
    \includegraphics[width=13cm]{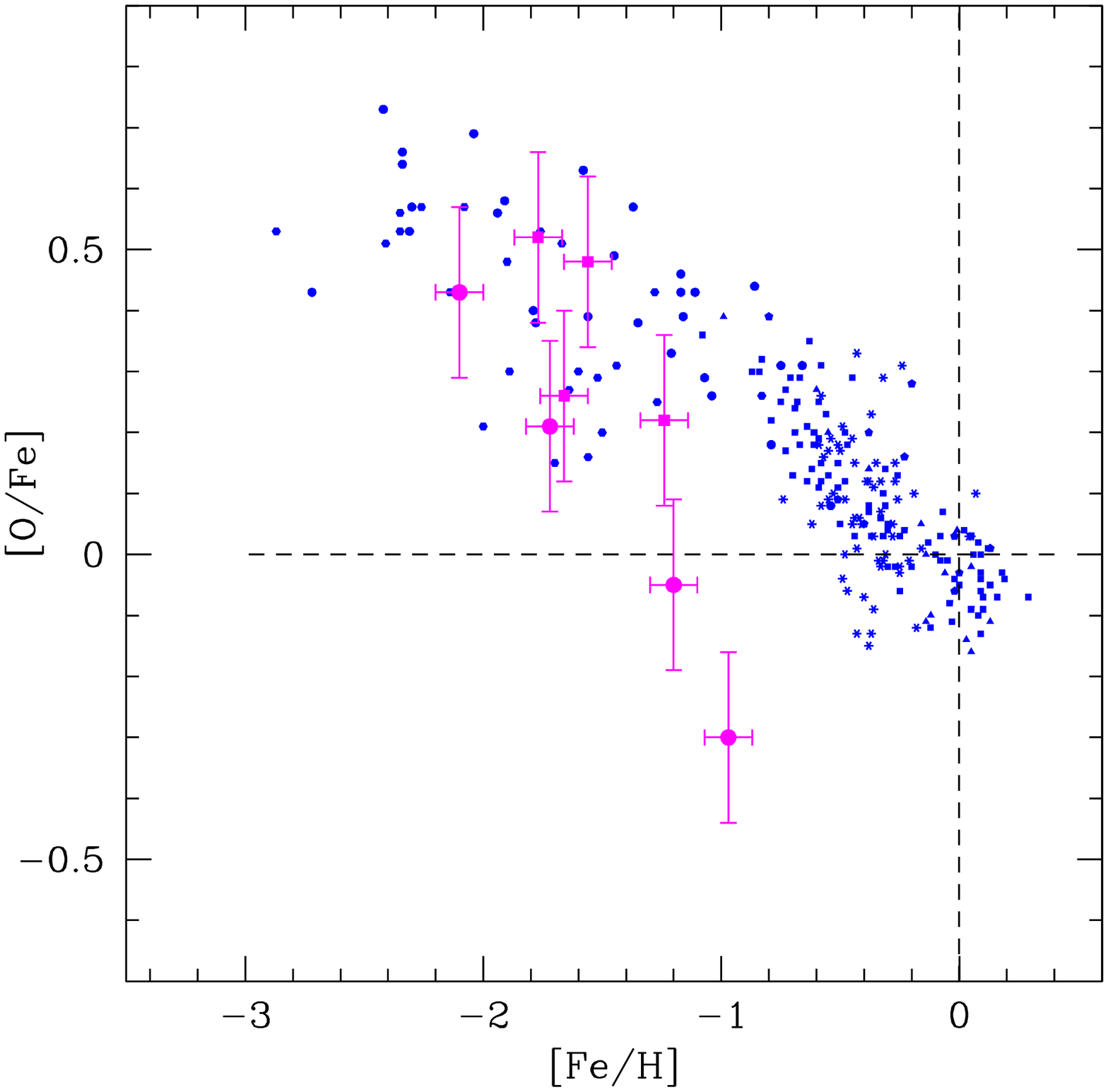}
\figcaption[fig3.ps]{
[O/Fe] versus [Fe/H] for the Sculptor red giants (large magenta 
filled circles - this paper, large magenta filled squares - S03), along with samples of Galactic field stars (small 
blue symbols).  The Galactic samples include only studies that use either
directly the [O I] 6300\AA\ line (and in some cases the 6363\AA\ line as well)
or the IR OH lines, or tie their results to the [O I] line.  
Due to the large numbers of Galactic
stars, their symbols must be kept small, so the different symbols are
not apparent, but these studies include
Edvardsson et al. (1993 - filled squares),
Cunha et al.  (1998 - filled pentagons), Melendez et al. (2001 -
filled hexagons), Smith et al. (2001 - filled triangles), Nissen
et al. (2002 - filled circles), or Reddy et al. (2003 - 6-pointed stars). 
}
\label{fig3}
\end{figure}

\begin{figure}
   \centering
    \includegraphics[width=14cm]{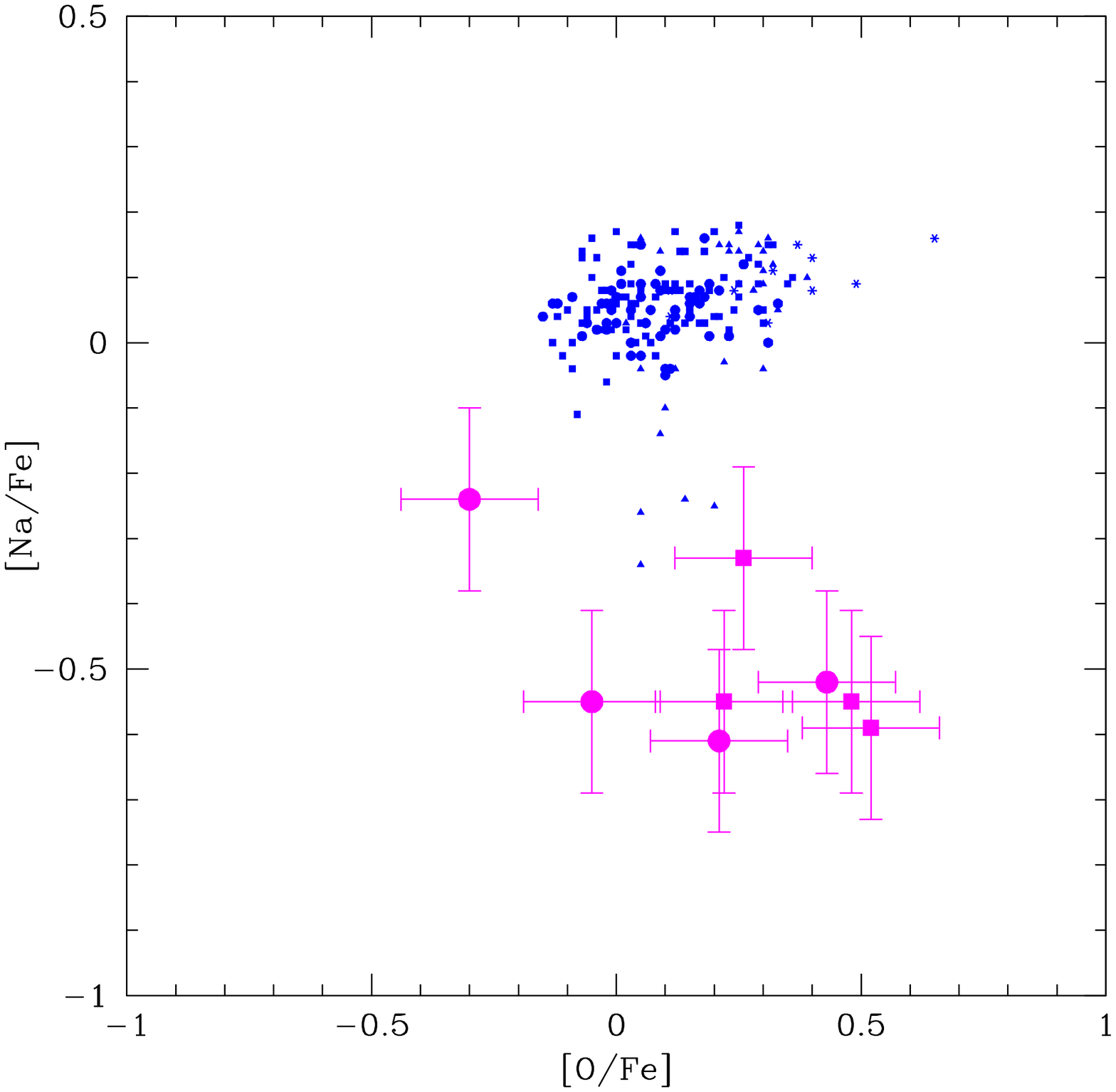}
\caption[fig4.ps]{
Values of [Na/Fe] versus [O/Fe] for the Sculptor red giants (large 
magenta filled circles - this paper, large magenta filled squares - S03), along with results from Galactic field stars (small
blue symbols). Due to the large numbers of Galactic
stars, their symbols must be kept small, so the different symbols are
not apparent, but these studies include
 Edvardsson et al. (1993 - filled squares), Nissen
et al. (1997 - filled triangles), Prochaska et al. (2000 - 6-pointed
stars), or Reddy et al. (2003 - filled circles).
}
\label{fig4}
\end{figure}

\begin{figure}
   \centering
    \includegraphics[width=16cm]{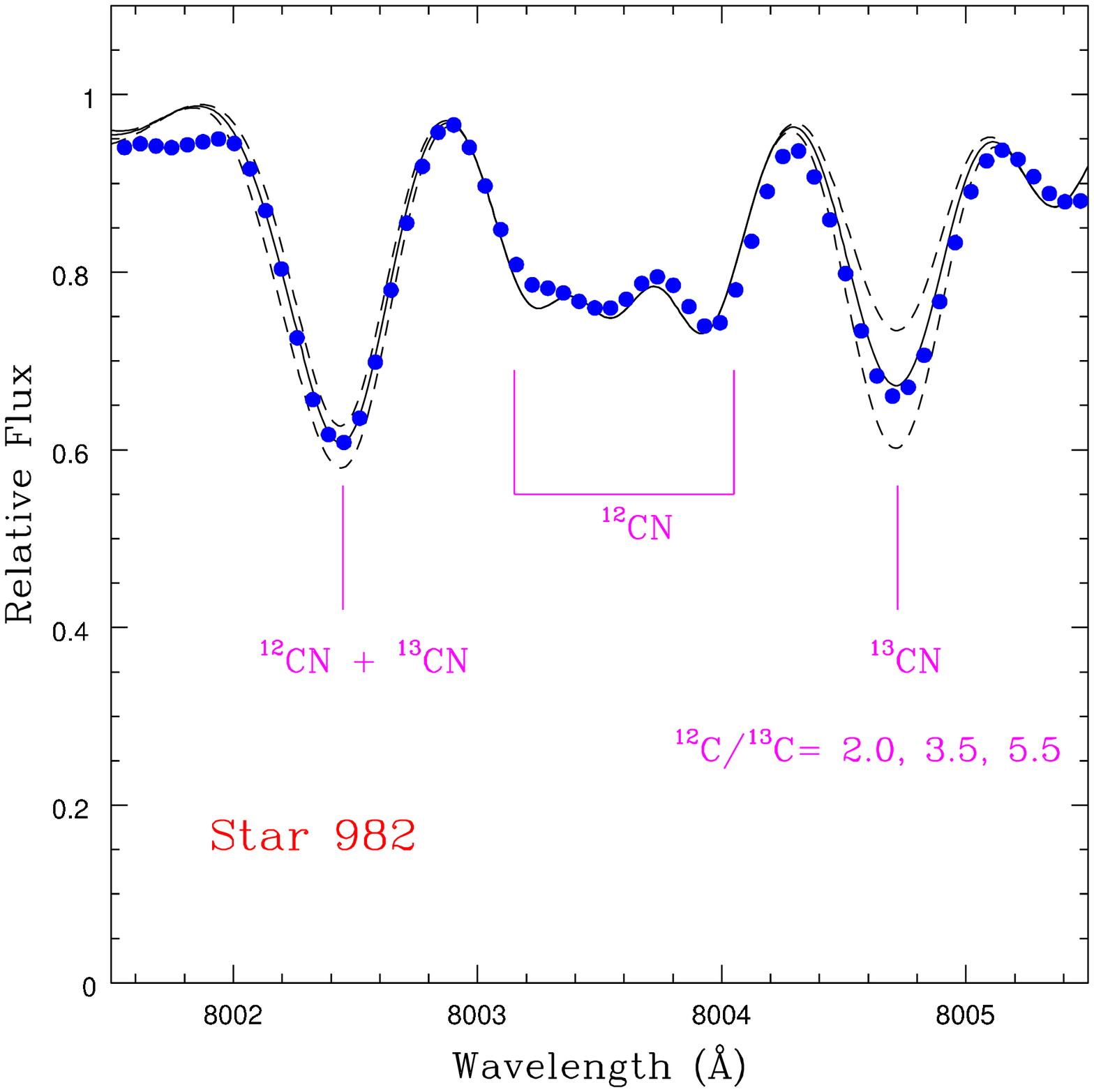}
\caption[fig5.ps]{
The CN lines near 8000\AA\ in star 982, as well as synthetic spectra
computed with three different $^{12}$C/$^{13}$C ratios.  Note that the
carbon isotope ratio in this red giant is about $^{12}$C/$^{13}$C= 3.5 --
close to the equlibrium ratio for the CN cycle.
}
\label{fig5}
\end{figure}

\begin{figure}
   \centering
    \includegraphics[width=14cm]{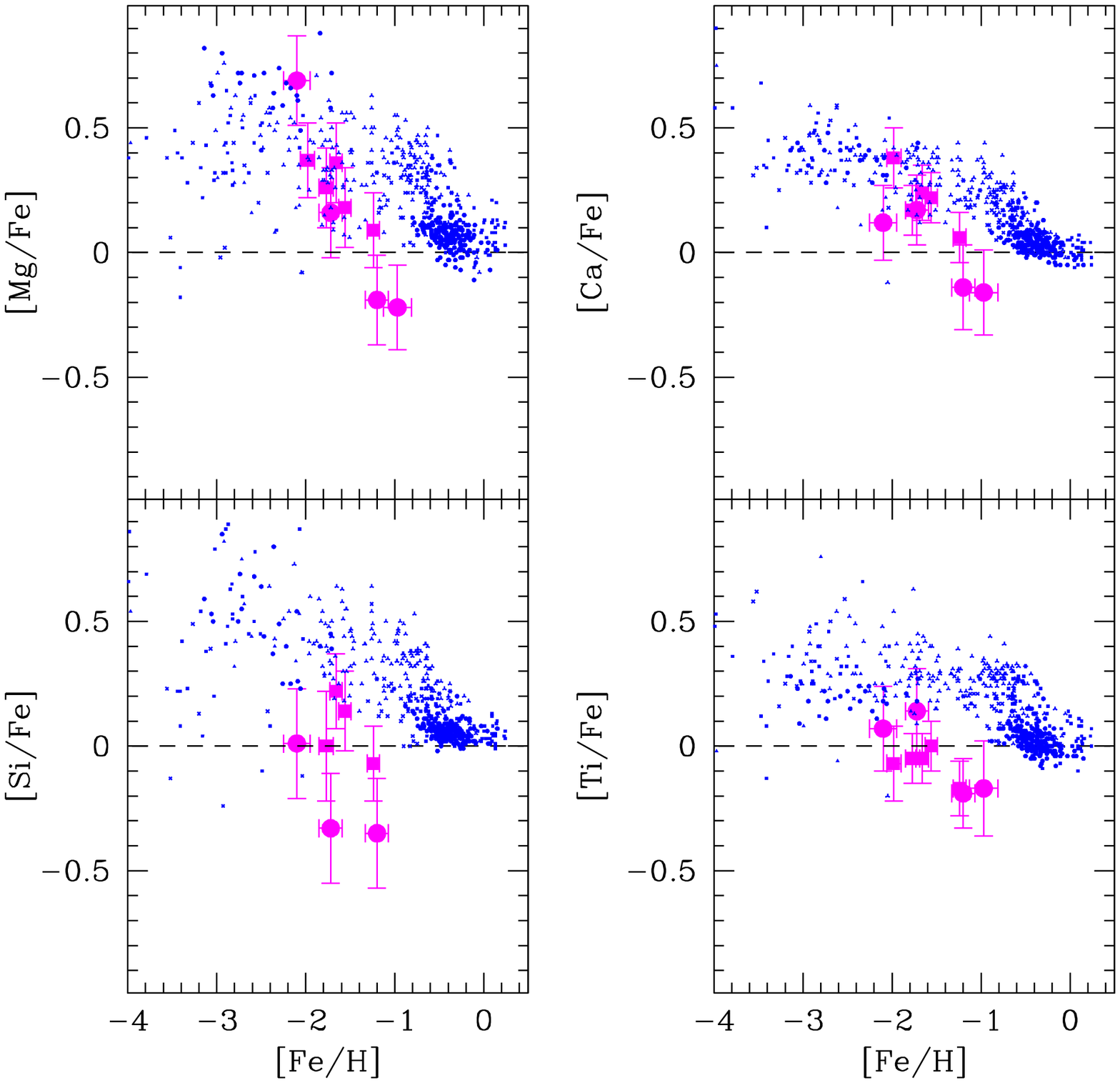}
\caption[fig6.ps]{
Values of [Mg/Fe], [Si/Fe], [Ca/Fe], and [Ti/Fe] versus [Fe/H]
for the Sculptor red giants (large magenta filled circles - this paper, large magenta filled squares - S03) and samples of Galactic
field stars (small blue symbols).  Due to the large numbers of Galactic
stars, their symbols must be kept small, so the different symbols are 
not apparent, but these studies include Gratton \& Sneden (1988),
Edvardsson et al. (1993), McWilliam et al. (1995), Nissen et al. (1997),
Prochaska et al. (2000), Carretta et al. (2002), Fulbright (2002),
Johnson (2002), and Reddy et al. (2003).  
}
\label{fig6}
\end{figure}

\begin{figure}
   \centering
    \includegraphics[width=14cm]{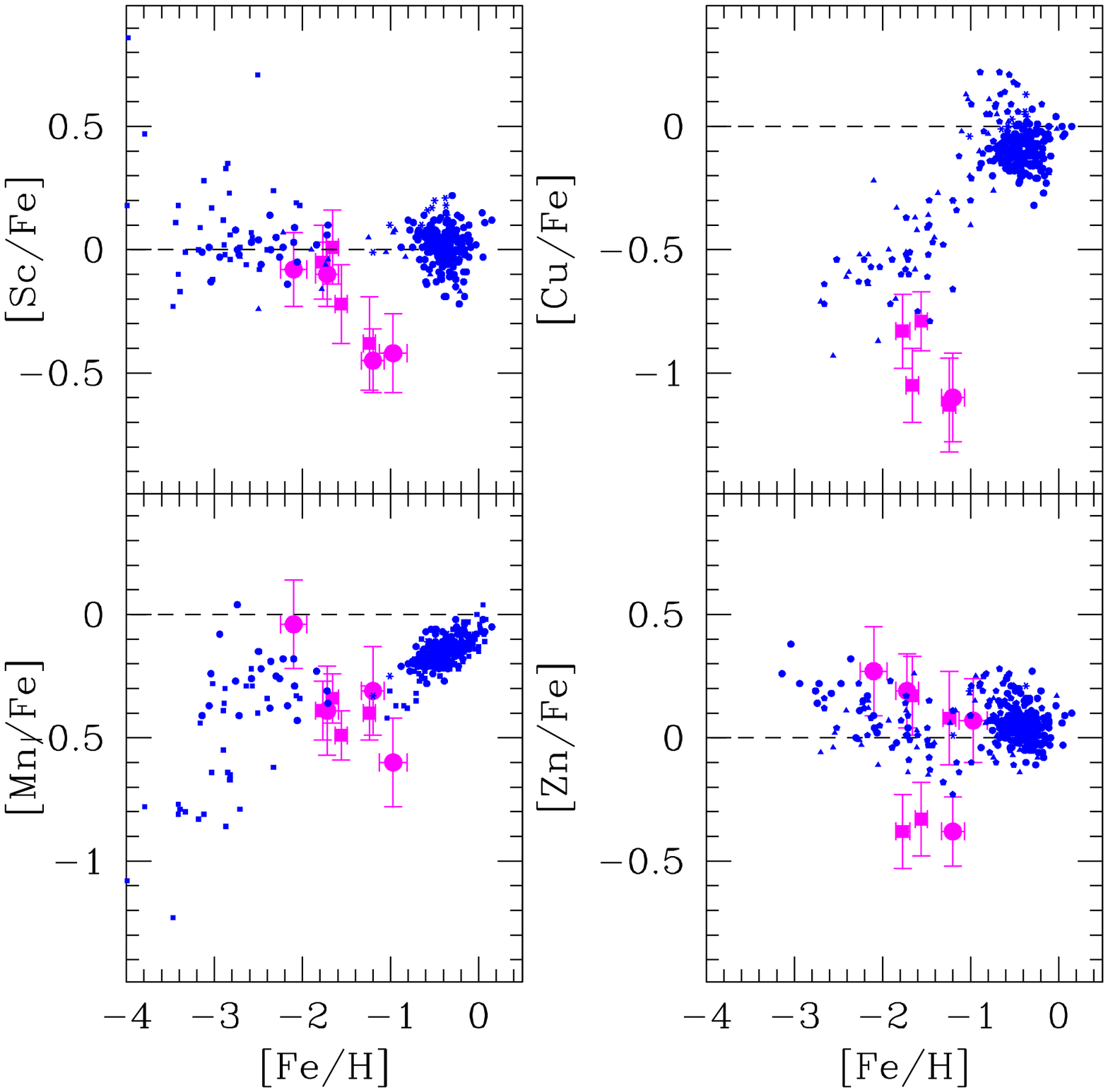}
\caption[fig7.ps]{
Values of [Sc/Fe], [Mn/Fe], [Cu/Fe] and [Zn/Fe] versus [Fe/H] for
the Sculptor red giants (large magenta 
filled circles - this paper, large magenta filled squares - S03) and samples of Galactic field stars (small blue 
symbols).  Due to the large number
of Galactic points, these symbols are kept small and the different
symbols are not apparent, but these studies include Sneden \&
Crocker (1988), Sneden et al. (1991), Gratton \& Sneden (1991), McWilliam
et al. (1995), Nissen et al. (2000), Mishenina et al. (2002),
Johnson (2002), and Reddy et al. (2003). 
} 
\label{fig7}
\end{figure}

\begin{figure}
   \centering
    \includegraphics[width=14cm]{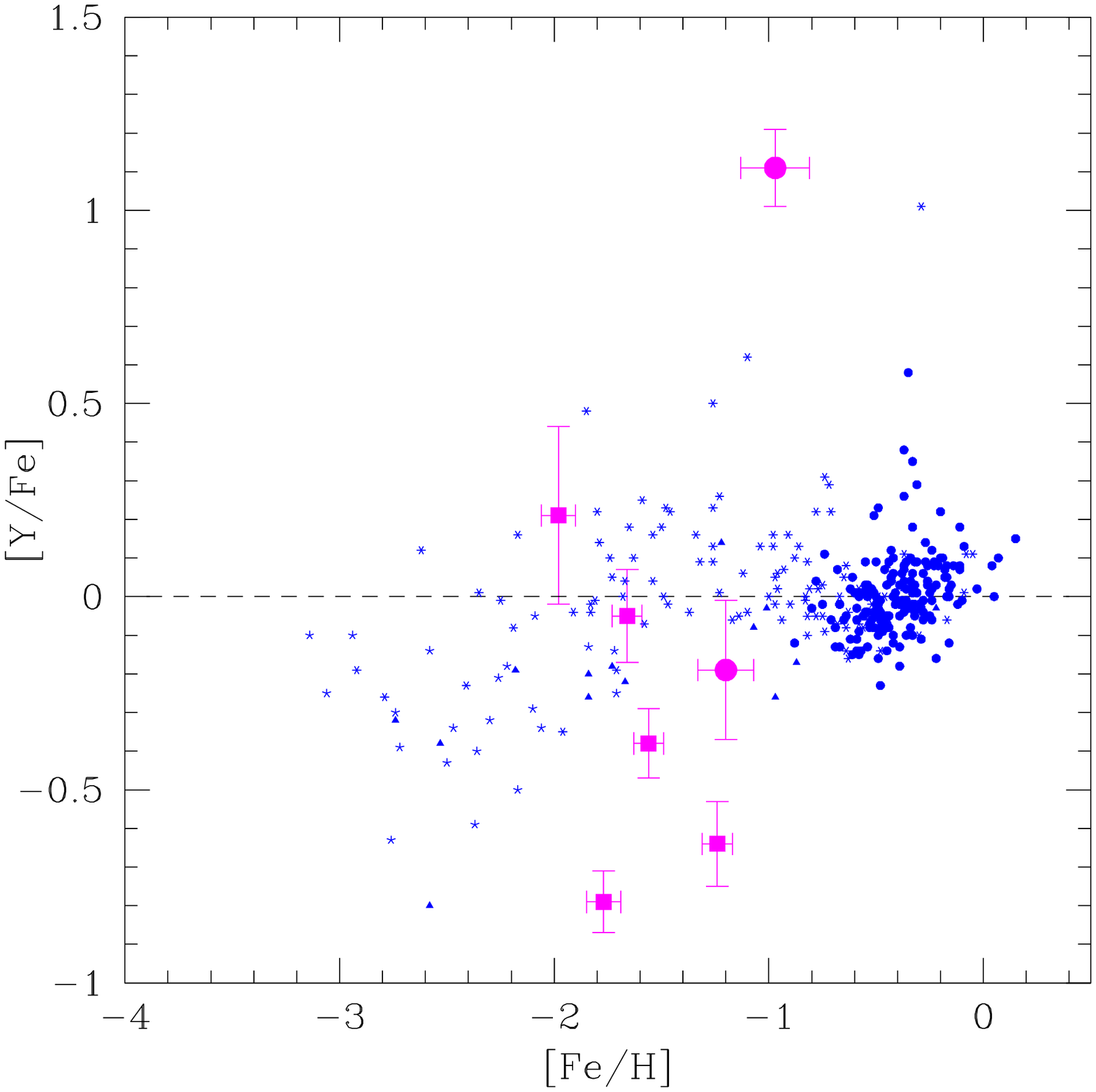}
\caption[fig8.ps]{
Values of [Y/Fe] versus [Fe/H] for the Sculptor red giants (large 
magenta filled circles - this paper, large magenta filled squares - S03) and samples of Galactic field stars (small 
blue symbols).
Yttrium is used as a surrogate for the light s-process elements (often
the average of Y and Zr are used, but most of the Galactic studies of
metal-poor stars do not include Zr; the addition of Zr to the Sculptor
stars would not change their positions significantly).  The Galactic
studies include Gratton \& Sneden (1994 - filled triangles), Fulbright
(2002 - 6-pointed stars), Johnson (2002 - 5-pointed stars), and
Reddy et al. (2003 - filled circles). 
Note the extreme [Y/Fe] enhancement in star 982.
}
\label{fig8}
\end{figure}

\begin{figure}
   \centering
    \includegraphics[width=16cm]{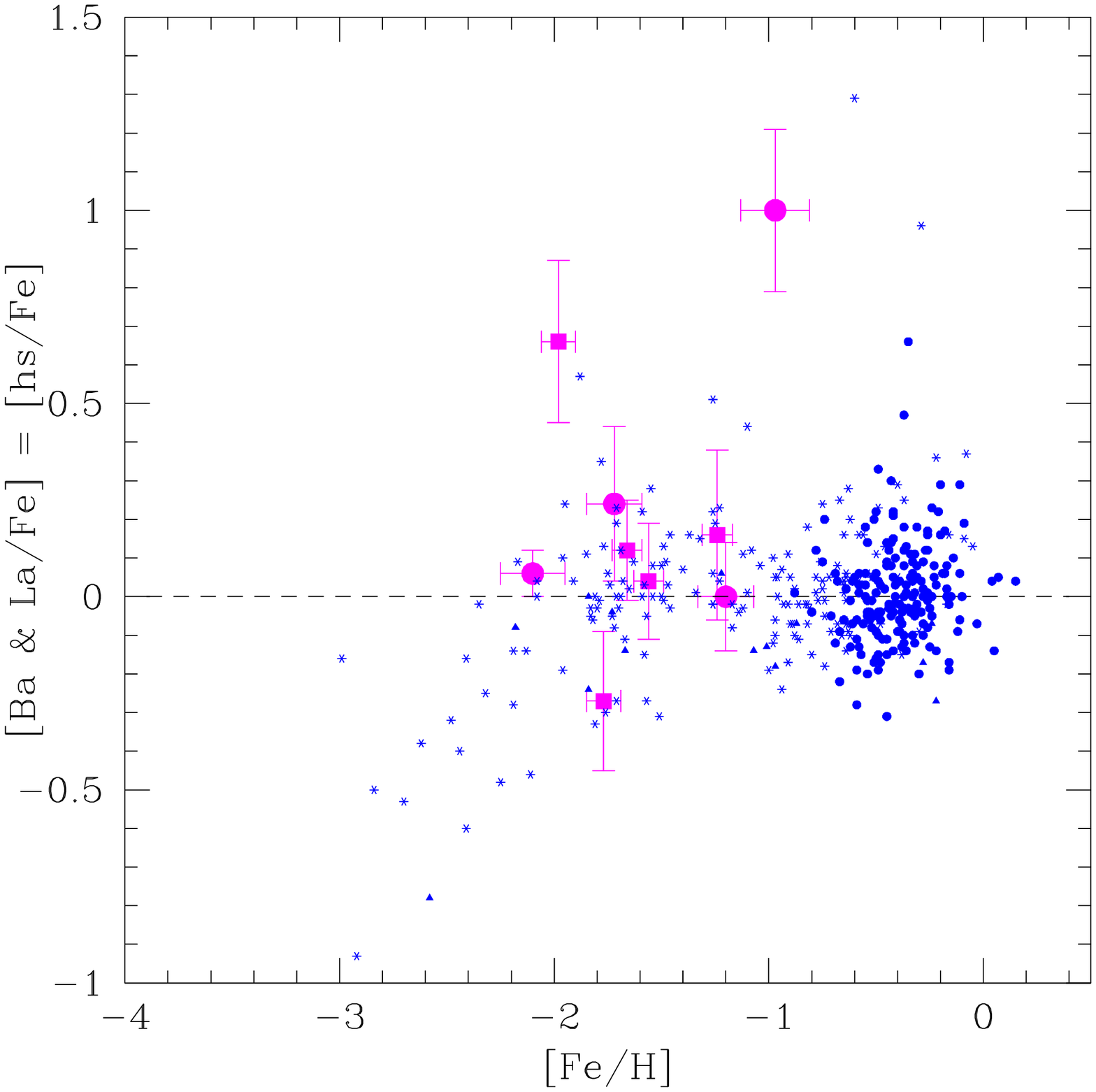}
\caption[fig9.ps]{
Values of [hs/Fe] versus [Fe/H], where ``hs" is the mean of the heavy 
s-process elements Ba and La, for the Sculptor red giants (large 
magenta filled circles - this paper, large magenta filled squares - S03) and Galactic field stars (small blue 
symbols). The Galactic studies consist of Gratton \& Sneden (1994 - filled triangles),
Fulbright (2002 - 6-pointed stars), and Reddy et al. (2003 - filled
circles). 
Note the extreme heavy s-process enhancement in star 982.
}
\label{fig9}
\end{figure}

\begin{figure}
   \centering
    \includegraphics[width=16cm]{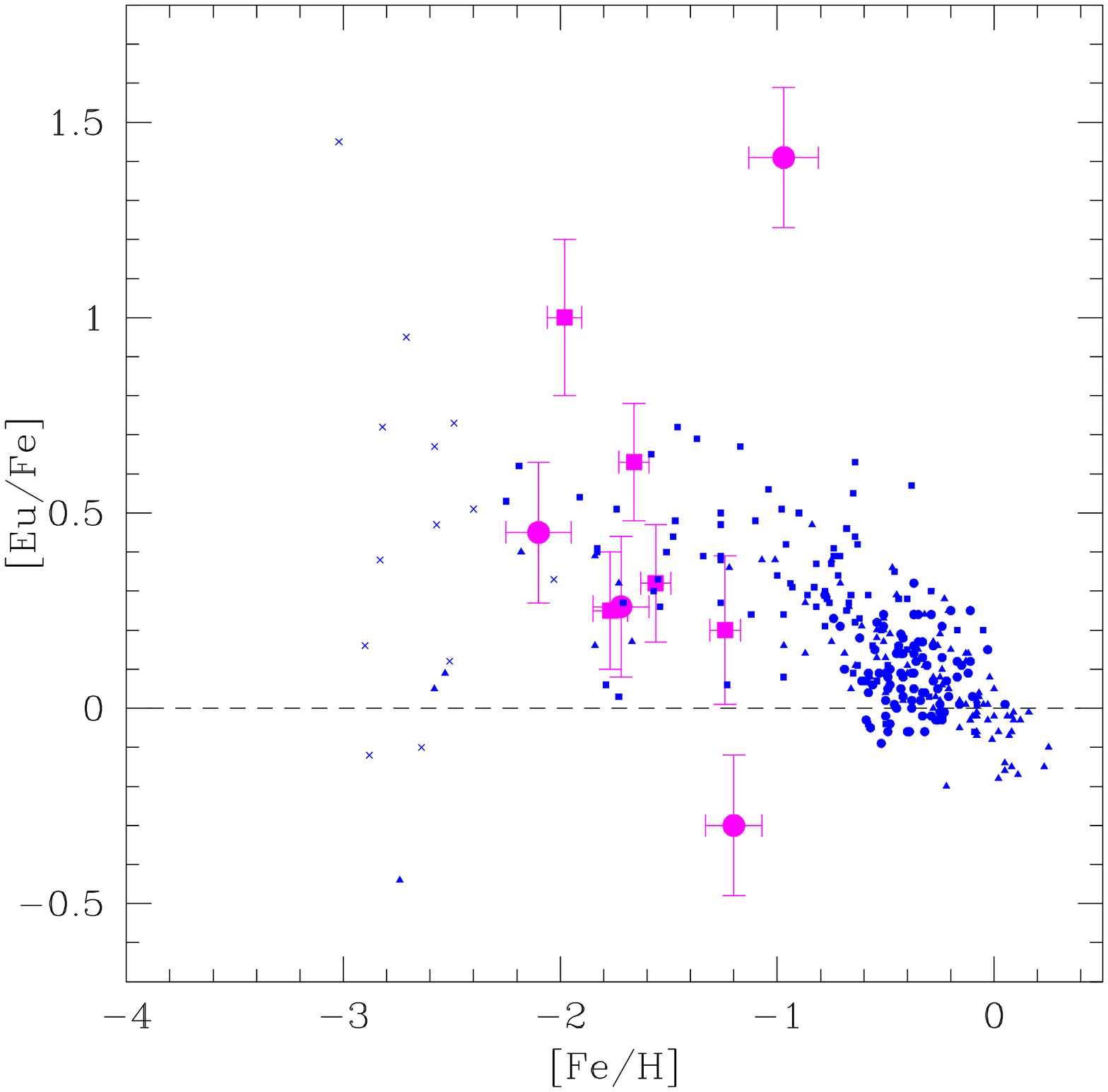}
\caption[fig10.ps]{
Values of [Eu/Fe] versus [Fe/H] for the Sculptor red giants (large 
magenta filled circles - this paper, large magenta filled squares - S03) and samples of Galactic field stars (small 
blue symbols). The Galactic studies include Gratton \& Sneden (1994 - filled triangles),
Woolf et al. (1995 - filled triangles), McWilliam et al. (1995 - 4-pointed
crosses), Fulbright (2002 - filled squares), Johnson (2002 - 6-pointed
stars), and Reddy et al. (2003 - filled circles). 
Note the extreme enhancement of star 982.
}
\label{fig10}
\end{figure}

\begin{figure}
   \centering
    \includegraphics[width=14cm]{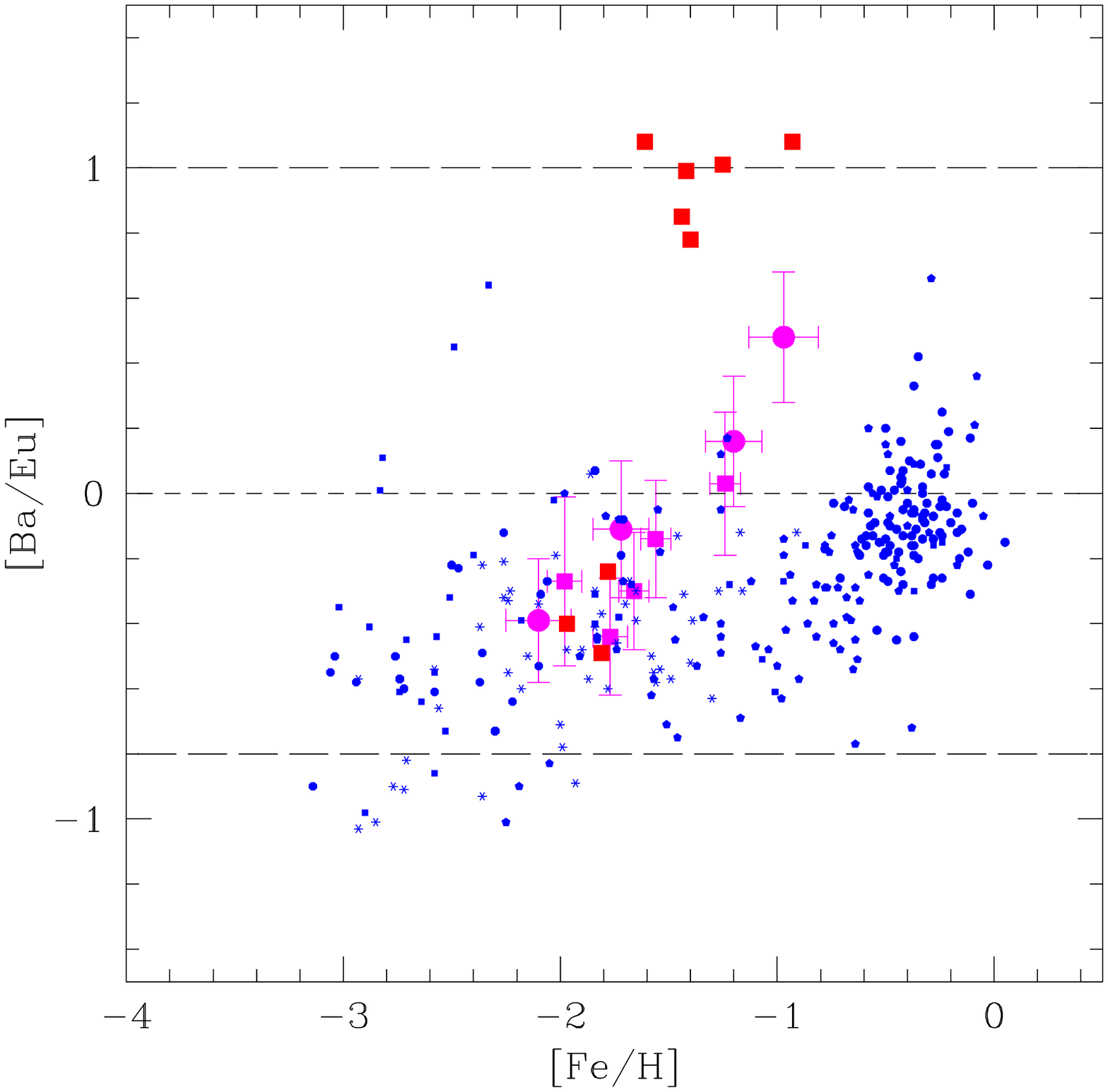}
\caption[fig11.ps]{
Values of [Ba/Eu] versus [Fe/H] for the Sculptor red giants (large 
magenta filled circles - this paper, large magenta filled squares - S03), Galactic field stars (small blue 
symbols) and red giants from the peculiar globular cluster $\omega$ Cen (large 
red squares). 
The long dashed line below represents the pure r-process abundance ratio 
and the long dashed line above represents the pure s-process abundance ratio 
from Arlandini et al. (1999).  The Galactic studies include Gratton \&
Sneden (1994 - filled squares), McWilliam et al. (1995 - filled squares),
Burris et al. (2000 - 6-pointed stars), Fulbright (2002 - filled
pentagons), Johnson (2002 - filled hexagons), and Reddy et al. (2003 -
filled circles).  The points for $\omega$ Cen are from Smith et al.
(2000).  
}
\label{fig11}
\end{figure}


\end{document}